 \documentclass[final,5p,times,twocolumn,authoryear]{elsarticle}

\usepackage{amssymb}
\usepackage{amsmath}
\usepackage{lipsum}

\usepackage{tikz}
\usepackage{braket}

\usepackage[breaklinks=true]{hyperref}
\usepackage{xurl}
\journal{Physics Open}

\begin{document}

\begin{frontmatter}



\title{ Multiple Nonlinear Waves by  (Quantum) Neural Networks: 
\\
Checking the AI supremacy }


\author[a,b]{L. Martina} 
\author[a]{R. Caricato}
\author[c,d]{R. Della Torre}
\affiliation[a]{organization={Università del Salento},Dipartimento di Matematica e Fisica
            addressline={}, 
            city={Lecce},
            postcode={73100}, 
            state={},
            country={Italy}}
\affiliation[b]{organization={INFN, Sezione di Lecce },
             addressline={}, 
            city={Lecce},
            postcode={73100}, 
            state={},
            country={Italy}}
\affiliation[c]{organization={IIT  Lecce },
            addressline={}, 
            city={Arnesano (LE)},
            postcode={73010}, 
            state={},
            country={Italy}}
\affiliation[d]{organization={Università del Salento},Dipartimento di Ingegneria dell'Innovazione
            addressline={}, 
            city={Lecce},
            postcode={73100}, 
            state={},
            country={Italy}}

\begin{abstract}
To test the potential of so-called quantum physics-informed neural network (QPINN) technology for solving evolutionary partial differential equations, we consider the problem of multiple wave propagation in a weakly nonlinear medium described by the Korteweg-de Vries equation with periodic boundary conditions. Although this problem is solvable analytically, high-genus  solutions could represent a rather challenging testbed for numerical methods. Therefore, our idea is to test whether a QPINN is sufficiently mature for application in this context by providing a set of indicators of merit and performing several numerical experiments.  
\end{abstract}




\end{frontmatter}




\section{Introduction}
\label{introduction}

In recent years, many deep Neural Networks focused on the study of nonlinear Evolution Differential Equations (PDEs), such as the Physics-Informed Neural Networks (PINN) \citep{raissi2017physics,RAISSI2019686,Raissi_Cuomo}.
This approach distances itself both from the traditional applications of Machine Learning (ML) techniques and from the much better-known analytical/numerical methods of PDEs. 
In fact, on the one hand, because of  the explosive growth of available data and computing resources, the ML has been
promoted  from  regression to classification problems and it has been
shown that deep neural networks can be used as general function approximator \citep{HORNIK1989359,SCARSELLI199815}.
While traditional machine learning can be trained from (possibly labeled) data without knowing the underlying mechanism that produced it, physical system modeling, by contrast, relies on general theoretical principles and (differential) equations for specific dynamical variables that are generally known, in some cases even long ago. Under certain approximations, the equations of motion can be expressed in linear form, whose solutions can be obtained with well-known methods. However, most differential equations are nonlinear and their solutions can exhibit chaotic behavior, irregular and sensitive dependence on initial conditions and, generally, cannot be expressed analytically. However, several nonlinear evolution problems have been solved with analytical and numerical methods, profoundly clarifying their geometric meaning. This is especially true for fully integrable systems \citep{ablowitz1981solitons,Ablowitz_Clarkson_1991,10.1093/oso/9780198872535.001.0001}, mainly in 1+1 and 2+1 dimensions. But a lot remains of  difficult cases to   effectively solve, like the celebrated (1+1)D Kuramoto-Sivashinsky equation \citep{HYMAN1986113}, the 3D Navier-Stokes equation \citep{Protas_Navier-Stokes}, the (N+1)D NonLinear Schroedinger \citep{Kenig_2006} and the analogous complex Ginzburg-Landau equation \citep{RevModPhys.74.99}, the 4D  Yang-Mills system \citep{BeinerTHEYE} and infinitely many other equations.  Moreover, also in the case of the completely solvable or integrable PDEs, perturbations and  boundary conditions may introduce  non integrable effects. So special studies are required in such a cases. Thus, new mathematical tools, including machine learning,  may be useful in investigating certain classes of solutions.  

In this context, in a first step \citep{raissi2017physics} a PINN has been proposed to solve evolutionary nonlinear partial differential equations. Then,  the soliton solutions of KdV, mKdV and KdV-Burgers equation were studied   by using the PINN
method \citep{Li_2020}. A little later  \citep{JAGTAP2020113028} the  PINN method was enriched by of local conservation laws  on discrete domain to solve Burgers and KdV equations. Then, the NLS equation was discussed in terms of the  PINN approach by \citep{Fang2021} and in particular rogue waves in the defocusing case was studied \citep{WANG2021127408}.

The main strategy of PINN is to minimize an appropriate loss function for the NPDE-related configurational variables, evaluated at the initial/boundary sampling points and on the evolution domain. In this way, the PINN method can provide feedback on the optimal parameters of the neural network. These requirements can be strengthened by imposing loss functions that involve any local NPDE conservation laws \citep{Raissi2020,LIN2022111053,FANG2022112118}.

The existence of conservation laws is particularly relevant in the case of fully integrable systems, as an infinite number of them can be found in commutation with each other. This aspect, however, has not yet been implemented. Furthermore, in such particular cases, Hamiltonian and bi-Hamiltonian structures emerge, opening up further possibilities for implementing numerical calculations and simulations, including PINN techiques. In any case, the key feature of integrable partial differential equations is that they are compatibility conditions for certain linear operators (Lax pairs). Analysis of the corresponding spectra leads to the identification of suitable separate variables, whose time evolution can be determined via quadratures. This allows for the writing of appropriate formulas for entire classes of global solutions. It is clear that in these situations, the machine learning approach offers no advantage in principle over analytical resolution, or its numerical transposition, unless there are significant difficulties in implementing the aforementioned formulas. These methods can be quite complex if they involve the calculation of numerous nonlinear modes (oscillations), which typically decompose into an exponentially large number of normal harmonic modes. Therefore, the first aspect we wish to explore in this work is whether the PINN approach is robust enough with respect to the complexity of the initial and boundary data, even for a well-known, fully integrable system. In particular, we chose to study the typical behavior of PINNs with respect to the propagation of multiple nonlinear waves (Stokes waves or cnoidal waves), described in the Korteweg-de Vries (KdV) model by higher-genus finite-gap solutions \citep{dubrovin1976non,B_A_Dubrovin_1981,belokolos1994algebro}.

Such a kind of problem is relevant in our context since: 1) the initial and boundary conditions are not trivial at all, 2) a formal analytic solution exists in terms of Riemann Theta functions, but it may involve an exponentially large number of harmonic components in order to get a numerically meaningful solution, 3) alternatively numerical solutions may be expressed in terms of a smaller, but sufficiently large, number of hyperelliptic functions \citep{frauendiener2004hyperelliptic,trogdon2013numerical} or by solving high order Riemann-Hilbert problems \citep{Bilman2022ComputationOL} , 4) a quite generic initial problem  for KdV may involve nonlinear superposition of solitons and cnoidal waves, 5) experimental data reported the nonlinear superposition  of hundred of nonlinear wave components \citep{osborne2010nonlinear}.  

For these studies, the analysis began using the standard PINN method. However, several different  routes are possible. Certainly, the one involving conservation laws is mandatory. On the other hand, modern machine learning techniques also hope to benefit from quantum computing, which promises to speed up the most complex computational tasks by exploiting the nonlocal correlations between entangled qubits. This perspective  is strongly supported by several studies \citep{schuld2018supervised,pastorello2023concise,du2025gentle}, which currently have relatively few hardware implementations, but which could be very interesting even in the current NISQ era \citep{Preskill2018quantumcomputingin}.

In particular, we compared the classical PINN method with a quantum version of the neural networks (QPINN)  based on Variational Quantum Algorithm (VQA)  \citep{benedetti2019parameterized,trahan2024quantum,PhysRevA.109.042421,schuld2018supervised}. 

The VQA were introduced for complex chemical calculations \citep{Peruzzo13}.  Applications such as simulating complicated quantum systems or solving large-scale linear algebra problems are very challenging for classical computers, owing to the extremely high computational cost. Quantum computing promises a solution, although fault-tolerant quantum computers will probably not be available in the near future. Current quantum devices have serious limitations, including low numbers of qubits and noise processes that limit circuit depth. VQAs have emerged as a leading strategy for addressing these constraints and use classical optimizers to control parameterized quantum circuits. VQAs have now been proposed for essentially all applications that researchers have envisaged for quantum computers, and they appear to be the best hope for obtaining quantum advantage. Nevertheless, challenges remain, including the trainability, accuracy and efficiency of VQAs. 

 In this work, we compared the capabilities of a QPINN with respect to a PINN in  the study of specific PDEs solutions, highlighting the fact that a reduction in the number of variational parameters is typically achieved.

Therefore, our main goals in this work can be summarized as follows:
1. To use (Q)PINNs in large systems, where standard numerical techniques become computationally expensive and unstable;
2. To provide a standard tool for comparing relative performance by training (Q)PINNs on fully integrable systems before applying them to more general and less structured evolutionary partial differential equations.

The article is organized, after this Introduction, with Section 2 summarizing the most relevant results from the theory of multiperiodic KdV solutions. Section 3 summarizes the main characteristics of a PINN algorithm and subsequently of a QPINN. Section 4 illustrates the details of the calculations performed on some examples of genus 2, 3, and 5 waves, using both PINN and QPINN.
Finally, Section 5 draws some conclusions from comparisons between the various examples reported and, more importantly, opens up new avenues for further investigation.

\section{Multi-periodic solutions of the KdV}
The celebrated Korteweg - de Vries equation (KdV) for the real field $u\left( x, t \right)$ reads \citep{ablowitz1981solitons} 
 \begin{equation}
     u_t-6uu_x+u_{xxx}=0 .\label{KdVeq}
 \end{equation}
 It is a universal model for   conservative dispersive non linear waves. For instance, it arises as a model for Non Linear Waves  in Shallow Water \citep{Whitham}, but it playes a role also in quantum gravity \citep{JT-gravity}.

 The key property for solving the KdV equation consists in the so-called \textit{Lax representation} \citep{ablowitz1981solitons}:
 \begin{eqnarray}
   &\hat L_t=[\hat L,\hat A],\\
  & \hat L=-\partial_x^2+u(x,t)\,, \qquad \hat{A}=4\partial_x^3-6u\,\partial_x-3u_x.
 \end{eqnarray}
The  Schr\"odinger operator $\hat{L}$ over ${\cal L}^2_{\mathbb R}$ allows to perform the mapping  
\begin{center}KdV data $\stackrel{IST}{\longleftrightarrow }  \hat L$ spectral data ,\end{center} where $IST$ understands for \textit{Inverse Spectral Transform}. Precisely, in solving the problem with regular initial conditions with rapidly vanishing potentials $x \, u\left( x, t\right) \stackrel{|x|\to\infty}{\longrightarrow} 0$, the following chain of three linear problems
has to be performed:  
\begin{tikzpicture}
    \node[shape=rectangle,draw=black] (B) at (-2,3) {{$u\left( x, 0 \right)$}};
    \node[shape=rectangle,draw=black] (C) at (4,3) {\small$
\begin{array}{rcl}
\sigma(\hat{L})
& = &
\left\{ i \sqrt{\kappa_n} \right\}_{n=1,\dots,N}
\;\cup\; \mathbb{R}^+ \\[0.5em]
\left\{ \beta_n \right\}
& , &
\rho(k_R)
\end{array}
$
}
;
    \path [->,>=stealth](B) edge node[above] {Direct  Prob.} (C);
   \node[shape=rectangle,draw=black] (D) at (4,1) 
   {\small{$
\begin{array}{rcl}
    \sigma\left( \hat{L} \right)&=&  \left\{ i \, \sqrt{\kappa_n}\right\}  \bigcup \mathbb{R}  \\
 \left\{ \beta_n \, e^{8  \kappa_n^{3/2} t} \right\}  &, &  \rho\left( k_R \right) \, e^{ 8 \, i \, k_R^3 \,t}    
\end{array}$}};
    \node[shape=rectangle,draw=black] (E) at (-2,1) {{$u\left( x, t  \right)$}};
      \path [->,>=stealth](D) edge node[above] {Inverse  Prob.} (E);
     \path [->,>=stealth](C) edge node[right] {\small{ Time Evolution}} (D);
  \end{tikzpicture}
Here the \textit{Direct Prob.} understands  the Schroedinger operator scattering problem  for the  potential  $u\left( 0, t \right)$, described in the momentum complex plane $k = k_R + i\; k_I$. While by \textit{Inverse  Prob.} we indicate the solution of the Gelfand-Marchenko equation \citep{ablowitz1981solitons}. For reflectionless potentials  ($\rho\left( k_R \right) \equiv 0$), one obtains   the well-kown   $N$-soliton Hirota formula 
\begin{eqnarray}
   & u =& -2 \partial_x^2\, \log\left[\det[A\left( x,t\right]\right],\label{Nsolitons}
    \\
    A_{n m} &=& \delta_{n m} + \beta_n\, \frac{\,
    e^{8 \kappa_n^3\,t -\left(\kappa_n + \kappa_m  \right) x}}{\kappa_n+\kappa_m},\nonumber \\ \; \small   k_n  &>0, \quad &\beta_n \in \mathbb{R}_{/0} , \quad n = 1, \dots , N \nonumber . 
\end{eqnarray} 
     The computational complexity of such a formula is basically related to that one of the computation of the determinants, that is $O\left( N^3\right) $, but instabilities can occur in the asymptotic regions $|t|\to \infty$ and $|x|\to \infty$. On the other hand, solving the initial data problem may results intriguing, mainly to prove reflectionless property of the potential. However, about such a topic there exists a wide literature \citep{ablowitz1981solitons,Matveev1992DarbouxTA}. 

     On the other hand, one can pose the question of periodic potentials  $u\left( x + L, t\right) = u\left( x, t\right) $. This problem was approached by the classical Floquet theory for the time-independent Schr\"odinger equation \citep{dubrovin1976non,B_A_Dubrovin_1981,belokolos1994algebro}.

Summarizing the main results, one looks at the   matrix eigenfunction solution
    \begin{equation}{ \mathbf{\Phi}}\left( x, x_0, k\right)  = 
\left(
\begin{array}{cc}
\phi  & \phi_x     \\
\phi^*  & \phi^*_x  
\end{array}
\right) 
\end{equation} 
of the Schroedinger equation with the \textit{initial} periodic potential $u\left( x , t=0\right) = u\left( x \right) $ and "energy" eigenvalue $E$, namely 
  \begin{equation}
     \mathbf{\Phi}_{x x } + \left[  u\left( x \right) +k^2\right]  {\mathbf{\Phi}} = 0 ,  \; \; k^2 = E  .
\label{Sch}  \end{equation}   

The matrix $\mathbf{\Phi}$ is  normalized at the point $x_0$ as follows 
 \begin{equation} { \mathbf{\Phi}} \left( x_0, x_0, k\right)= 
\left(
\begin{array}{cc}
1  & i \, k    \\
1  & - i \, k  
\end{array}
\right) .  \end{equation}
  
  The periodicity of the potential implies the existence of a \textit{monodromy matrix} $ \mathbf {T}\left( x_0, k \right) $ such that
  \begin{equation}
      { \mathbf{\Phi}}\left( x + L, x_0, k \right) =  \mathbf {T}\left( x_0, k \right)  { \mathbf{\Phi}}\left( x, x_0, k \right),
  \end{equation}   
 which, in its turn,  defines the so-called  main spectrum 
 \begin{equation}
     {   \left\{ E_j = k_j^2: \; |\; \frac{1}{2} \textrm{{ Tr}}\left[ \mathbf T\left( E_j \right)\right]| =  1 \right\}_{1 \leq j \leq 2N+1} }. \label{MSpect}
 \end{equation}   
    The set of gaps  $\left\{\left.\right] -\infty, E_1\left.\right], \left[ E_{2 j} , E_{2 j +1 }\right] \; {1\leq j \leq N}\right\}$ determines  $N+1$ forbidden zones for the existence of bounded regular solution to the system (\ref{Sch}). If $N < \infty$, $ u\left( x \right) $ is called a finite-gap potential of \textit{genus} $N$,  and these are of greater interest to physics. For values of $E$ outside the gaps, the   scalar wavefunctions $\phi \left( x, x_0, E\right),\; \phi^* \left( x, x_0, E\right) $  are meromorphic on the two-sheeted  Riemann surface $  \Gamma$ covering  the $E$-plane 
    \begin{equation}
       W^2=P_{2N+1}(E)=\prod_{i=1}^{2N+1}(E-E_i)\, .
   \label{Gamma} \end{equation}
The wavefunctions $\phi \left( x, x_0, E\right),\; \phi^* \left( x, x_0, E\right) $    have  a simple pole in the complex $E$-plane,  located in the band-gaps at $\gamma_j(x) \in[E_{2j},E_{2j+1}]_{1\leq j\leq N}$.  The ends of the gap zones $E \to E_i$  are branch-points. The following relations hold
\begin{eqnarray}
    & \phi \, \phi^*  = \prod_{i = 1}^N \frac{E- \gamma_i\left( x\right)}{E- \gamma_i\left( x_0\right)},& \nonumber\\
   &  -\frac{i}{2} \, W\left[ \phi,  \phi^*\right] = \frac{ \sqrt{\prod_{i = 1}^{2N+1} \left( E- E_i \right)}}{ \prod_{i = 1}^N \left( E- \gamma_i\left( x\right)\right)} 
 \stackrel{k \to \infty}{\rightsquigarrow} k + \sum_{n \geq 0} \frac{\chi_{2 +1} \left( x \right)}{\left( 2 k\right)^{2+1}},\label{wronskian}
\end{eqnarray}
where$W\left[ \cdot ,  \cdot \right]$ denotes the Wronskian of its arguments and  $\gamma_i\left( x\right)$ are  the so called hyperelliptic functions. Their values $\gamma_i\left( x_0 \right)$  and the  sheet where  the poles  are located complete the entire
set of spectral data, together with the main spectrum (\ref{MSpect}).

The asymptotic series expansion in inverse power of $k$, in right hand side of expression (\ref{wronskian}), allows to determine completely the evolution on time $t$ of the functions $\gamma_i$. Thus, one is led to the following compatible system 
  \begin{equation}
  u(x,t)=-2\sum_{j=1}^N \gamma_j(x,t)+\sum_{j=1}^{2N+1}E_j ,
           \quad     \alpha(\gamma_j) =-2(u+2E)|_{E=\gamma_j(x)}\,.\label{hyperallisol}\end{equation}
  \begin{equation}
                \gamma_j'(x, t)= \pm\,\frac{2i\sqrt{P_{2N+1}(\gamma_j)}}{\prod_{k\neq j}(\gamma_k-\gamma_j)}\,,\quad    
                 \dot \gamma_j(x,t) = \pm\,\frac{2i\alpha(\gamma_j)\sqrt{P_{2N+1}(\gamma_j)}}{\prod_{k\neq j}(\gamma_k-\gamma_j)}\, ,\label{dubrovincurve}
\end{equation}
where $  \gamma_j'(x, t)$ and $\dot \gamma_j(x,t) $
    represent $x-$ and $t-$derivative, respectively. 
 
In the above formulation, the Inverse Spectral Transform for periodic potentials is given by the solution of the two  separated ODEs systems presented in (\ref{dubrovincurve}), sometimes called the Dubrovin curve.  For $N=1$ the equation for $\gamma_1(x, t )$ correspond to uniform translation of a suitable  Weierstrasse elliptic function. But already for $N=2$ the ODE system is challenging to be explicitely solved \citep{dubrovin1976non} and numerical techniques are required for higher $N$ \citep{trogdon2013numerical,Bilman2022ComputationOL}.

On the other hand, Dubrovin  discovered an implicit way to linearize   the problem (\ref{dubrovincurve}) in general, by using the algebraic geometry.
The first step is to introduce on the elliptic curve (\ref{Gamma})  a family of  non contractible oriented curves encircling  the gap bands, or going from one sheet to the other one by crossig the gap band. They form a homotopy group and a base of such curves are cycles denoted by  $(a_i,b_i)_{i=1,\ldots,N}$.  A dual description is provided by   a canonical bases of holomorphic  differentials  on $\Gamma$
\begin{equation}
   d\omega_i \left( E \right) = \sum_{j=1}^N 
   c_{i, j} \frac{E^{j-1}}{\sqrt{P_{2N+1}\left(E\right)}}dE \quad \;{i = 1, \dots , N} , \end{equation} normalized in such a way  
\begin{equation}
     \int_{a_j}d\omega_i= \delta_{j i}\; .
\end{equation}
Thus one can introduce the symmetric matrix of periods $\mathbf B \in\mathbb{C}^{N \times N}$ with  entries
     \begin{equation}
          B_{ij}=\int_{b_j}d\omega_i =  \sum_{m=1}^N c_{i m }  \int_{E_1}^{E_{2j}}  \frac{E^{m-1}}{P^{1/2}_{2N+1}\left( E\right)} dE 
     \end{equation}
The matrix $\mathbf B$ has negative real part $\Re{\mathbf B} < 0$.
Thus, the matrix $\mathbf B$ allows  to build the Riemann Theta function associated to the Rimannian surface $\Gamma$, namely 
\begin{eqnarray}
     \Theta_N( \mathbf z|\;\mathbf B )=
     \sum_{\mathbf n\in \mathbb{Z}^N}
    e^{\,\frac{1}{2}\,\mathbf n \cdot \mathbf B\,\cdot\,\mathbf n + i \mathbf n\,\cdot\,\mathbf z } \,
    \; ,  \mathbf z =\mathbf U\,x - \mathbf W\,t+\mathbf \Psi\,\in\mathbb{C}^N, \label{RTheta}\end{eqnarray}
   where $\mathbf U$, $\mathbf W$ and $\mathbf \Psi$ are suitable constant $N$-component vectors belonging to $\mathbb{C}^N$.
    
    The function $\Theta_N$ is quasi-N-multiperiodic, in the sense that 
  the relation \begin{eqnarray} \Theta_N(\mathbf z +2 \pi i \mathbf N + \mathbf B\cdot \mathbf M|\;\mathbf B)= e^{-\frac{1}{2} \mathbf M \cdot \mathbf B \cdot \mathbf M - i   \mathbf M\cdot \mathbf z } \; \Theta(\mathbf z|\;\mathbf B)
 \end{eqnarray}
 holds for any arbitrary $N$-components vectors of integers  
 $ \mathbf M$ and $ \mathbf N$.
 
Then, the main result is that a $N$-multi-periodic  solution of the KdV equation (\ref{KdVeq}) is expressed by 
\begin{equation}
    u\left( x, t \right) = -2 \partial^2_x \log  \Theta_N\left( \mathbf z|\;\mathbf B \right), \label{solHiroTheta}
\end{equation}
where the components of the $\mathbf z$ variable introduced in \eqref{RTheta}  are specified  by the so-called Abel map
\begin{equation}
    z_j = - \imath \sum_{m=1}^N \int_{E_{2m}}^{\gamma_m\left( x, t \right)} d \omega_j = U_j x - W_j t + \Psi_j 
\end{equation}
in  terms of the spectral data. 
This is a fundamental relationship between the solution in terms of hyperelliptic functions 
(\ref{hyperallisol}) and  the one  (\ref{solHiroTheta}) in terms of the Riemann $\Theta$, or in Fourier expansion form, if it preferred.
Both formulations are very useful, but each of them has computational limitations. 
From one side, the computation of the hyperelliptic functions is challenging, on the other hand the Fourier components to sum up increases exponentially with the genus N. 
\begin{figure*}
    \centering
    \includegraphics[width=.7\linewidth]{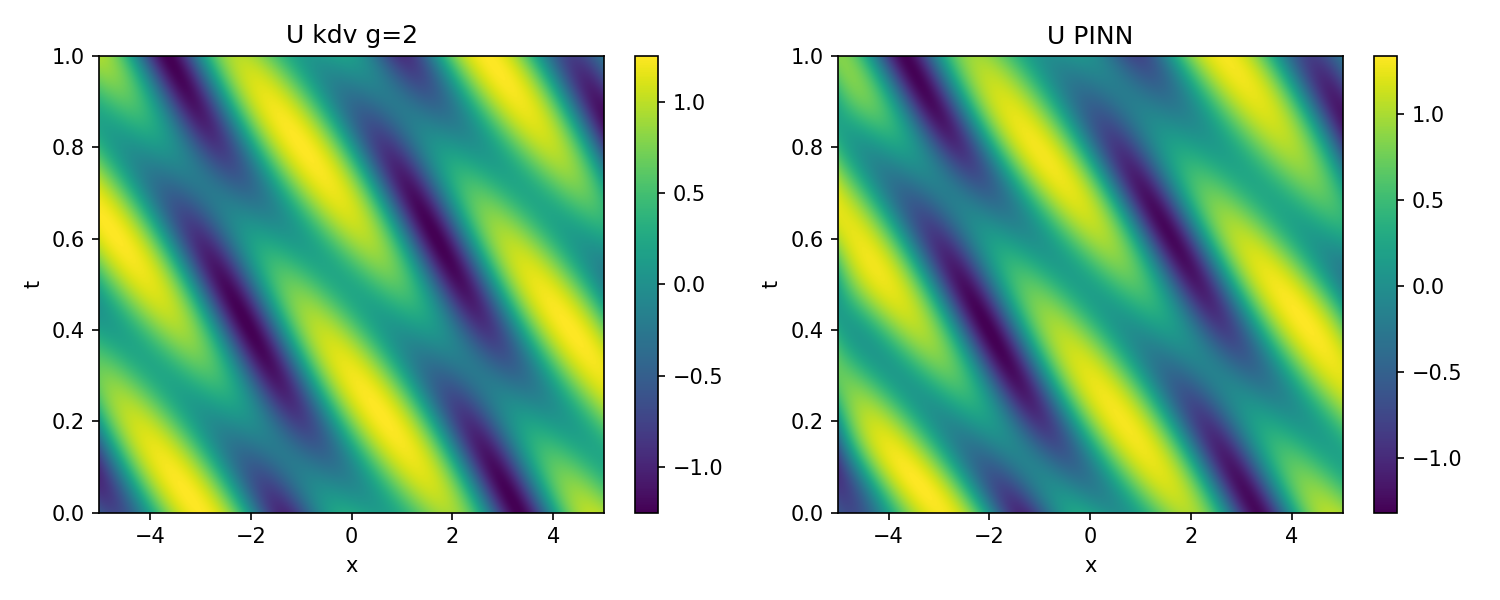}
    \caption{$g=2$ solution of \eqref{KdVeq} Comparing the numerical solution (left) with the   PINN calculations (right)}
    \label{fig:enter-label1}
\end{figure*}
In order to be more explicit,  the Riemann $\Theta_N$ function of genus $N$  can be written as 
\begin{eqnarray} &\Theta_N \left( x,t \right)= &\sum_l\, q_le^{i\left( K_l x - \Omega_l t + \Psi_{ l}\right)}, \;\; 
q_l= e^{- \left( 1/2 \right)\mathbf n_l\cdot \mathbf B \cdot \mathbf n_l} ,  \;  \\
 &K_l= {\mathbf n}_l\cdot {\mathbf k},& \quad  { \Omega}_l= {\mathbf n}_l\cdot {\boldsymbol \omega}, \quad { \Psi}_l= {\mathbf n}_l\cdot {\mathbf\Psi}, \;\; \\   &  \mathbf n_l \in \mathbb{Z}^N ,& \;{\mathbf k} \in \frac{2 \pi}{L} \mathbb{Z}^N , \;  \quad \omega_j =  k_j^3 \sigma\left(k_j, \mathbf B\right)\Big|_ {j = 1, \dots , N}\label{dispersion}\end{eqnarray}
 where the last relation in (\ref{dispersion})  is specific of the KdV equation. So, even if the wave-vectors ${\mathbf k}$ are commensurables, the same does not holds for the frequency vectors $\boldsymbol\omega$.

 In the special case $N=1$, the Riemann $\Theta_1$ reduces to the Jacobi $\theta_3$ function. In fact, taking the definition (\ref{RTheta}) defined as 
 \begin{eqnarray}
     \Theta_1\left( z , t\right) = \sum_{l = -\infty}^\infty \exp\left\{ i\, l \, k \, z -\frac{1}{2}\, b\, l^2 \right\} \stackrel{\Re\left(b\right) >0}{=} \nonumber \\
   =  1+ 2 \sum_{l=1}^\infty q^{l^2} \cos\left( l\, k\, z \right) = \theta_3\left( k\,z \,|\; q\right) , \;\; q =   e^{- \pi \, b} ,
 \end{eqnarray}
 where $$z =   x - k^2 \frac{4 \left(2m - 1\right) K\left[  m \right]^2}{\pi^2}  t . $$
Setting $b =  \frac{K\left[ 1-m \right]}{K\left[ m \right]} $ with $K\left[ m \right]$ denoting the complete elliptic integral of first kind of module $m$, one derives the KdV cnoidal solution 
\begin{eqnarray}
&u_{cn}\left(x,t \right) =  &- 2\,\partial_{x}^2\log\theta_3\left( k\, z  | q \right) = \\
& - 2 k^2 \frac{K\left[ m \right]^2}{\pi^2}& \left[  m \,\textrm{cn}^2\left(  \frac{K\left[ m \right]}{\pi} k \, z \, |\,  m \right)  + 1-m  - E\left[ m \right]/K\left[ m \right] \right] .\nonumber\label{cnoidalW}
\end{eqnarray}
Remarkably, the linear dispersive waves   are the small amplitude limit 
$$\lim_{m\to 0} u_{cn}\left(x,t \right)  \rightsquigarrow -4k^2 m \cos\left( k \left(x +  k^2 t\right)\right).$$

On the other hand, the limit $m\to 1$ of the expression \eqref{cnoidalW} will lead to the 1-solitonic solution by the Hirota formula (\ref{Nsolitons}). Of course, the nonlinear one mode is the simplest and a very special case. 

In principle, from an initial data point, one should derive the entire Riemann spectrum, i.e., the principal spectrum $\{ E_i\}$, i.e., the period matrix $\mathbf B$, and the phases.
However, in general, this requires the use of several advanced numerical techniques. Note that in several concrete circumstances (see, for example, \citep{osborne2010nonlinear}), hundreds of nonlinear modes overlap and interact. In typical experimental situations, the amount of computational resources required becomes an issue.

\section{Briefs about standard numerical approaches for higher genus solutions}
A first approach  in numerical integration of the periodic KdV equation for given initial and boundary data consists in solving the system (\ref{hyperallisol}-\ref{dubrovincurve}) in the following steps.

\begin{enumerate} 
\item  One provides a space/time discretization of the domain  $(x, t)\in[-L , L ]\times[0, T]$  by  a lattice of $S_x \times S_t$ points.
\item Chose the genus value $N$ 
\item  At $t=0$, the initial data are given by  the band edges of the  hyperelliptic surface: $\{E_j\}_{j=1,\ldots,N }$ and
      by  the initial conditions  $\{\gamma_j(x_0=0,0)= \gamma_{j,0}\}_{j=1, \dots , N}$ for the hyperelliptic functions at the  specific point  $x_0=0$ .
     \item   The integration procedure begins providing $\gamma_j(x_n, 0)$  by solving the Cauchy problem 
     \begin{eqnarray}
       \gamma_j'(x, 0)=\,\frac{\pm 2i\sqrt{P_{2N+1}(\gamma_j\left( x, 0 \right))}}{\prod_{k\neq j}(\gamma_k\left( x, 0 \right)-\gamma_j\left( x, 0 \right))},  \;  \gamma_j(0,0)= \gamma_{j,0} .  \nonumber
     \end{eqnarray}
\item At any space lattice site $x_n =  n  L/S_x$,   perform the time integration  of 
\begin{eqnarray}\hspace{-1.0cm}
     \dot \gamma_j(x_n,t)&=& \pm\,\frac{4 i \left( u\left( x_n, t \right) + 2 \gamma_j(x_n,t)\right)
     \sqrt{P_{2N+1}(\gamma_j\left( x_n, t \right))}}{\prod_{k\neq j}(\gamma_k\left( x_n, t \right)-\gamma_j\left( x_n, t \right))}\,,\nonumber
     \nonumber
\end{eqnarray}
 for $t\in[0,t_1=T/S_t]$ from the initial datum  $\gamma_j(x_n,0)$  
\item Update the value of the KdV solution  $$u(x_n , t_1)=-2\sum_{j=1}^N \gamma_j(x_n , t_1)+\sum_{j=1}^{2N+1}E_j$$ 
on the discretized lattice.
\item Repeat the procedure from 5. while $t_{S_t} <  T$, then print $u(x_n , T)$.   
\end{enumerate}
The above equation where numerically studied in 
\citep{osborne2010nonlinear,frauendiener2004hyperelliptic}. In \citep{trogdon2013numerical,Bilman2022ComputationOL} a differet approach was adopted.



\section{PINN and QPINN methodology}
Rewriting   \eqref{KdVeq} in the generic form of an evolutive nonlinear 1-space dimensional PDE 
\begin{equation}
      u_t(x,t)+\mathcal{N}[u](x,t)=0\,,\quad(x,t)\in \Omega\times[0,T]\,. \label{generica eq.ne evolutiva}
\end{equation}
where $\mathcal{N}[u]$ represents a nonlinear differential polynomial of the dependent variable $u(x,t)$ w.r.t. $x$. For sake of simplicity here we consider only constant coefficients w.r.t. the independent variables $x, t$. The previous equation is supplemented 
with initial and boundary conditions:
\begin{align}
    &u(x,0) = h(x)\,, \quad x \in \Omega\,, \\
    &\mathcal{B}[u;\, g_\pm(t)] = 0\,, \quad t \in [0,T]\,, \quad x =\partial\Omega\,.
\end{align}In this notation $\mathcal{B}$ is a boundary operator (for instance the Dirichlet, or the Neumann boundary conditions, but also mixed or other types are possible).

 The left hand side in \eqref{generica eq.ne evolutiva} is a function in the jet space $J^l$ of the variables $\left( x, t, u \right) $, which order $l$ equals  the highest derivative appearing in  $\mathcal{N}[u]$. Thus, it is natural to introduce   for any indexed family of elements $\left( x, t,  u_\theta,  u_{\theta\; t},  u_{\theta\; x}, \dots ,  u_{\theta\; t^l}, u_{\theta\; x^l},   \right)\in J^l $ the  residual function 
\begin{equation}
    \mathcal{R}_{\theta}(x,t)= u_{\theta\; t}(x,t)+\mathcal{N}\left(u_{\theta}(x,t), \dots, u_{\theta\;  x^l}(x,t)\right)\,.
\end{equation}

The best approximation $u_{\theta}(x,t)$ for the solution $u(x,t)$ of \eqref{KdVeq}  can obtained by minimizing the Loss function
\begin{equation} \label{Loss}
      \mathcal{L}_\theta(X)
= \mathcal{L}_\theta^{\mathcal{R}}(X^\mathcal{R})+  \mathcal{L}_\theta^{\text{ic}}(X^\text{ic})+  \mathcal{L}_\theta^{\text{bc}}(X^\text{bc})\,,
\end{equation}
over   the set of data $X$ used for the supervised training. The index $\theta$ here  is  understood as the set of trainable network parameters  optimizing the approximation.

Explicitly writing out the individual terms of the Loss function  one has: 
\begin{enumerate}
    \item Mean square error of residual
    \begin{equation}
    \mathcal{L}_\theta^{\mathcal{R}}(X)=\frac{1}{N_r}\sum_{i=1}^{N_r}|\mathcal{R}_{\theta}(x^i,t^i)|^2
\end{equation}
    \item Mean square error for initial condition
    \begin{equation}
    \mathcal{L}_\theta^{\text{ic}}(X)=\frac{1}{N_{\text{ic}}}\sum_{i=1}^{N_{\text{ic}}}|u_{\theta}(x_{\text{ic}}^i,0)-h(x_{\text{ic}}^i)|^2
\end{equation}
    \item Mean square error for boundary condition (both Dirichlet  and Neumann) 
    \begin{align}
    \mathcal{L}_\theta^{\text{bc}}(X)=\frac{1}{N_{\text{bc}}}\sum_{i=1}^{N_{\text{bc}}}|u_{\theta}(x_{\text{bc}\pm},t_{\text{bc}}^i))-g_{\pm}(t_{\text{bc}}^i)|^2+\\  
    \frac{1}{N_{\text{bc}}}\sum_{i=1}^{N_{\text{bc}}}|\partial_x u_{\theta}(x_{\text{bc}\pm},t_{\text{bc}}^i))-\partial_x g_{\pm}(t_{\text{bc}}^i)|^2
\end{align}
\end{enumerate}

\subsection{Details about PINN calculations}

The architecture used in this work is based on a  relatively simple deep
feed-forward neural networks architectures with hyperbolic tangent activation functions.   The feedforward neural network called Multilayer Perceptron (MLP)  is a parametrized  approximator  of functions 
\begin{equation}
    \mathbf{x}=(x,t) \to  \mathbf{f}_{\theta}(\mathbf{x})
\end{equation}  
where, in our calculations, the  tensor type input data is passed to a series of $ L = 7$ hidden layers with $40$ neurons for each layer.

Setting $\mathbf{r}^{(0)}(\mathbf{x)}=\mathbf{x}$,  the MLP is recursively defined by 
\begin{equation}
    \mathbf{f}^{(l)}_{\theta}(x)=\mathbf{W}^{(l)}\cdot \mathbf{r}^{l-1}+ \mathbf{b}^{(l)}\, ,\quad \mathbf{r}^{(l)}(x)=\sigma(\mathbf{f}_{\theta}(x)),\;\; l=1,2\ldots,L.
\end{equation}

Then,  the final output layer yields
\begin{equation}
    \mathbf{f}_{\theta}(x)=\mathbf{W}^{(L+1)}\cdot \mathbf{r}^{(L)}(x)+\mathbf{b}^{(L+1)}\,.
\end{equation}
In the adopted notation  above  $\mathbf{W}^{(l)} \in\mathbb{R}^{m_l\times d_{m-1}}$ is the weight matrix in l-th layer, thus collectively    $\theta = (\mathbf{W}^{(1)}, \mathbf{b}^{(1)},\ldots, \mathbf{W}^{(L+1)}, \mathbf{b}^{(L+1)})$ represents all trainable parameters in the network.

Furthermore, the symbol $\sigma$ denotes an element-wise activation function, which for our network is the hyperbolic tangent (Tanh).

 Minimization of $\mathcal{L}_\theta$ is made by  performing an iterative  gradient descent algorithm. Where, at the  $n$-th iteration, the trainable parameter $\theta_{\mu}$ is updated according to 
\[
\theta_{\mu}(n{+}1)=\theta_{\mu}(n)-\eta\,
\left.\frac{\partial\mathcal{L}_\theta}{\partial \theta_{\mu}}\right|_{\theta=\theta(n)} ,
\]
being $\eta$  the \emph{learning rate} and
the   $\nabla_{\theta}\mathcal{L}_  \theta$ gradients  are calculated by using the \textit{Backward  Automatic  Differentiation} method. Such a kind of algorithm is called Adam \citep{kingma2014adam}.

\subsection{Details about QPINN calculations}

The architecture of the Quantum PINN  is a modification of the  classical one,   where   one hidden layer is replaced by a quantum node, simulated by resorting to the Pennylane library \citep{Ville}. The quantum node  is structured in a series of three different tools: i) the quantum data encoding of classical inputs (the quantum feature map), ii) a  unitary $U_{\theta}$ quantum variational layer composed by a set
of single-qubit parametric rotations alternated  by a corresponding set of  qubits entanglement operators, iii) a projective measurement apparatum. In our architecture      we  used $8$ variational layers  of $ 5$ qubit embedded  in the classical layers. 

\begin{figure}
    \centering
    \includegraphics[width=0.5\linewidth]{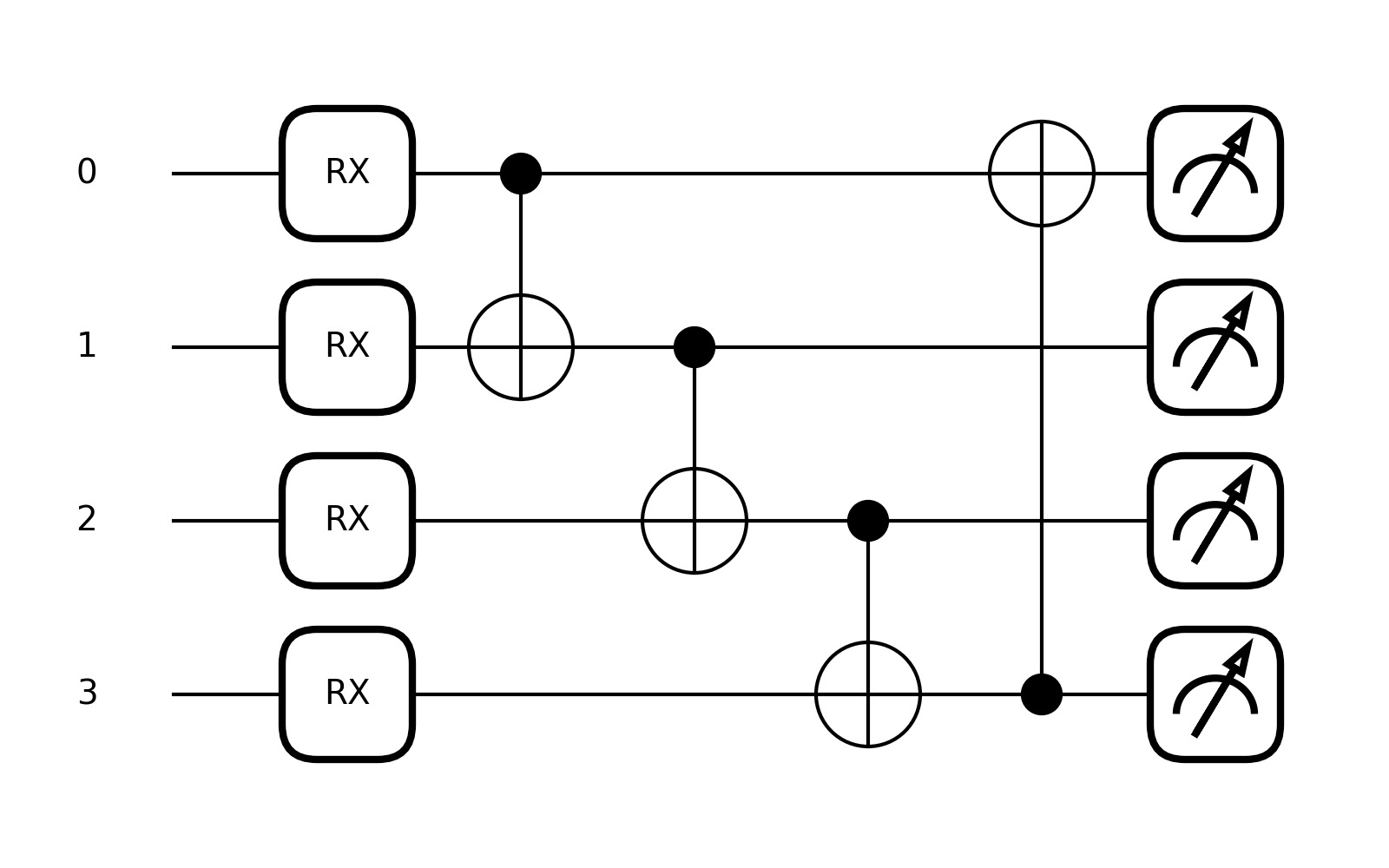}
    \caption{Graphical rappresentation of the \textit{BasicEntanglerLayers}}
    \label{fig:enter-label2}
\end{figure}

The quantum variational layer tunable parameters are the $\theta$ angles in the $R_X(\theta)$ rotations. The entanglement is realized with the CNOT gate between nearby qubits in a closed chain. A prototypical example of a such type is drawn in Fig.\ref{fig:enter-label2}  
but other architectures are possible.

 To produce a quantum parameterized output to add to the other "classical" $\theta$ parameters of the network, the qnode workflow is

 \begin{enumerate}
    \item Initialize the quantum  state
    \begin{equation*}
       \ket{\psi_0}=  \ket{0}^{\otimes n}\in \mathbb{C}^{2^n}
    \end{equation*}

    \item  Traslate the classical data $\mathbf{x}\in \mathcal{D}=\mathbb{R}^\text{m}$, being   $m$ the number of  input features, by  a suitable Angle Map $\mathbf{x} \to \mathbf{\phi} \in \left[ 0, 2\pi\right.\left[\right. $ and generate the quantum state
    \begin{equation*}
        \ket{\psi_0}=\ket{0}^{\otimes n}\rightarrow 
        \ket{\psi_{\text{enc}}}=U_{\phi}(\mathbf{x})\ket{0}^{\otimes n}
    \end{equation*}

    \setcounter{enumi}{2} 
    \item Apply the variational circuit $U_{\theta}$ providing 
    \begin{equation*}
        \ket{\psi(\mathbf{x};\theta)}=U_{\theta}U_{\phi}(\mathbf{x})\ket{0}^{\otimes n}
    \end{equation*}

    \item Measure the observable $\hat M \rightarrow \{M_k\in\sigma(\hat M),p_k(\mathbf{x};\theta)\}$ 
  
    \item Output from the network ; $\hat y=\phi^{-1}(<\hat{M}>)$.
\end{enumerate}
The simulations performed in the present work used the Pennylane functions  \textit{AngleEmbedding}, to yield the $\ket{\psi_{\text{enc}}}$ state, and  \textit{BasicEntanglerLayers}  in order to entangle the  qubits.

Since currently does not exists an efficient quantum algorithm for computing derivatives, the $\theta$ parameters introduced by the qnode will enter in the Loss function  \eqref{Loss} at the same foot of those coming from the classical layers. This leads to the  idea of the hybrid training of variational algorithms, in which queries to a quantum device are   optimized  by a classical algorithm.
The updated parameters are fed back to the quantum hardware in a closed loop, defining at each iteration  a new quantum
circuit.

The simulations performed in the present work do not take account of quantum noise effects.   

\section{Results Analysis}
\begin{figure*}
    \centering
\includegraphics[width=.7\linewidth]{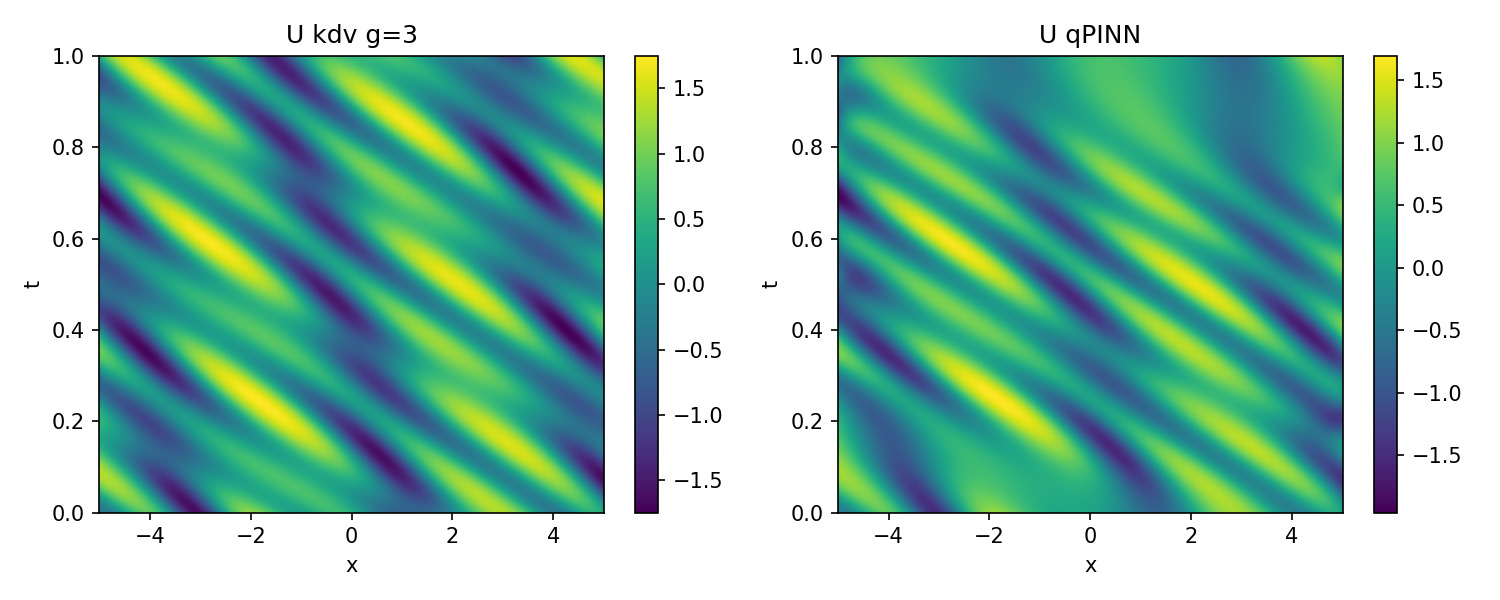}
    \caption{$g=3$ solution of \eqref{KdVeq}: comparing the numerical solution (left) with the   QPINN calculations (right)}
    \label{fig:enter-label3}
\end{figure*}
For the simulations we have used a time domain $t\in[0,1]$ and a space domain $x\in[-5,5]$. The numerical solutions were obtained by using standard integration methods for the Dubrovin's curve \eqref{dubrovincurve}.     But, for convenience, the input data were normalized inside the neural network, both for space $x$ and time $t$  to be in the range $\left[-1,1\right]$  by the relation
\begin{equation}
    x_n=2\frac{x-x_{\text{min}}}{x_{\text{max}}-x_{min}}-1\,, \qquad x_n\in \Omega\,.
\end{equation}

The training consist in $5000$ Adams epochs, followed by  $1600$ LBFGS  iterations. The latter belongs to the
family of second-order quasi-Newtonian methods, i.e., based on the calculation
of the Hessian matrix \citep{liu1989limited}.

Below the table of  the sampling points used to train the neural network:

\begin{tabular}{|l|l|}
\hline
\multicolumn{2}{|c|}{\textbf{Sampling points}}\\
\hline
PDE 
   & $\{(x_f^{(i)},t_f^{(i)})\}_{i=1}^{N_f}$,\ $N_f=8000$ \\
IC ($t=0$) 
   & $\{x_{ic}^{(j)}\}_{j=1}^{N_{ic}}$,\ $N_{ic}=800$ \\
BC value in $x=-5$ 
   & $\{t_{L}^{(k)}\}_{k=1}^{N_{bct}}$,\ $N_{bct}=800$ \\
BC value in $x=5$ 
   & $\{t_{R}^{(k)}\}_{k=1}^{N_{bct}}$,\ $N_{bct}=800$ \\
BC deriv. $x=-5$ 
   & $\{t_{Lx}^{(k)}\}_{k=1}^{N_{bct}}$,\ $N_{bct}=800$ \\
BC deriv. $x=5$ 
   & $\{t_{Rx}^{(k)}\}_{k=1}^{N_{bct}}$,\ $N_{bct}=800$ \\
\hline
\end{tabular}

Generalization means to compute the values of the function $u_{\theta}(x,t)$ in points not belonging to the sampling points and compare them with the true values of the numerical solution of \eqref{KdVeq}. 
We proved this \textit{integration}  method for different but small genus configurations. Among several experiments, here we are going to show the results (see Fig.\ref{fig:enter-label1} and Fig.\ref{fig:enter-label3}) concerning  two particular choices, being meaningful representatives of total set:
\begin{enumerate}
    \item $g=2$ case:

Spectral endpoints $E_j=[0.0,0.25,1.0,1.5,2.0]$\\
Auxiliary spectrum $\gamma_j(0)=\{0.5, 1.75\}$

\item $g=3$
 
Spectral endpoints $E_j=[0.0,0.25,1.0,2.0,2.5,3.0,3.5]$.
Auxiliary spectrum $\gamma_j(0)=\{0.5, 2.2,3.2\}$

\end{enumerate}

For the PINN case with the aforementioned architecture, we summarize the final losses, the relative error with respect to the exact numerical solution $u_{\text{exact}}$  defined by  
    $\mathcal{L}_2= \frac{\left\lVert u_{ \theta}-u_{\text{exact}}\right\rVert_2}{\left\lVert u_{\text{exact}}\right\rVert_2}$    
 and the total number of trainable parameters in the following table: 

\medskip
\begin{center}
  {\begin{tabular}{|c|*{2}{c|}}
\hline
{genus} & 2 & 3 \\
\hline
Final Loss & 6,0597e-02 & 1.5073e+00  \\
$\mathcal{L}_2$ error & 1,286e-01 &  2.912e-01  \\
Parameters & 10.001 & 10.001  \\
\hline
\end{tabular}}  
\end{center}

\medskip

For the QPINN approach,  after replacing  one hidden layer with a quantum node for $5$ qubits and $8$ variational layers, the corresponding results are 
\begin{center}
   \begin{tabular}{|c|*{2}{c|}}
\hline
{genus} & 2 & 3 \\
\hline
Final Loss & 1.4359e-01 & 1,9714e+00  \\
$\mathcal{L}_2$ error & 1,342e-01 &  3.671e-01  \\
Parameters & 8.571  & 8.571  \\
\hline
\end{tabular} 
\end{center}

\medskip

Furthermore, we have studied the variations of the relative error $\mathcal{L}_2$ as a function of the number of layers ad neurons, both in the classical ad quantum approach. 

\section{Conclusions and Perpectives}
From the study reported above we can draw some preliminary conclusions.

We have shown that it is possible to use (Q)PINNs in systems with a certain degree of complexity, here represented by the genus $g$ of the periodic KdV solutions sought. However, at the stage of this work, $g$ is still small, while we are interested in its values at least in the tens if not hundreds, where standard numerical techniques quickly become computationally expensive and unstable. This somewhat fulfills the goal 1. set in the Introduction.
However,  for the proposed architectures of neural network the $\mathcal{L}_2$ typically fluctuates in the range $10\% \, - \, 30\%$. Thus, the accuracy of the computations is not yet well established. 

 Comparing the previous tables, we see that QPINNs can significantly reduce the number of network parameters. However, the quality in terms of the $\mathcal{L}_2$ metric is slightly lower than the corresponding PINN in our scheme. In any case the relation among  architecture - number of parameters - genus- relative error is far to be completely investigated and understood.   
 
 Another aspect not yet considered is the evaluation of the minimal sampling points number required by both (Q)PINN architectures in relation with a given $\mathcal{L}_2$.

We noticed also that the constructed (Q)PINNs are unstable at times greater than the training time domain.
Furthermore, the stability of the network with respect to variations in the initial and boundary data parameters ($u_0$, $u_{R/L}$, $u_{x, R/L}$ ) needs to be verified.

These are the basic problems to be solved in order to establish an effective advantage of the ML techniques  with respect the analytical numerical one in the context of the studies about the evolutive PDEs. Since in our research we are comparing the PINN  approach with analytical very well  controlled contexts, it is much harder to make definitive  statements in the case of non integrable equations. 

Analogously, it is still difficult to evaluate the effective advantages of the quantum ML, for a long series of arguments. First of all, during the training the computational speed of the quantum algorithm on a classical computer is quite slow in comparison with true quantum computations.

 On the other hand, currently the implementation of variational quantum circuits is feasible with the  NISQ quantum devices. Thus, the above problem can be directly evaluated and it is i the program of our future researches. 

 Differently, the implementation of the gradient descent algorithm on quantum hardware is still a hot research topic. Thus, 
 the algorithm's architecture will remain hybrid for a long time to come. 

Thus, several aspects in the  training of (Q)PINNs are largely to be investigated, so that tests  on fully integrable systems could provide a standard tool for comparing their relative performance,
before applying them to non-integrable cases. 

Concerning the completely integrable systems, in particular the KdV equation, a very promising direction of investigation is to use the formulation of the Inverse Problem in terms of Riemann-Hilbert problem on the complex plane of the spectral parameter. This approach was implemented in numerical algorithms \citep{trogdon2013numerical,Bilman2022ComputationOL} able to manage solutions with higher $g$ than previously. From point of view of the  (Q)PINN approach this is a true challenge in terms    of effectiveness of the method. This is the topic which we are currently studying.   

Finally, another direction of investigation is to use (Q)PINN to compute spectral data for very complex cases observed in Nature. Indeed, this mathematical problem is analogous to, but much more extensive than, the Fourier transform method for linear PDEs. However, to our knowledge, there is currently no conventional quantum algorithm, such as the Quantum Fourier Transform, that significantly speeds up these calculations. Therefore, an alternative method could be a good contender.

\section*{Acknowledgements}
ML thanks the MMNLP project of the INFN for partially supporting this project.  




\bibliographystyle{elsarticle-harv} 
\bibliography{bibliografia.bib}






\end{document}